# Single-crystal epitaxial europium iron garnet films with strain-induced perpendicular magnetic anisotropy – structural, strain, magnetic, and spin transport properties

M. X. Guo,[1,#] C. K. Cheng,[2,#] Y. C. Liu,[1,#] C. N. Wu,[3] W. N. Chen,[1] T. Y Chen,[4] C. T. Wu,[5] C. H. Hsu,[6] S. Q. Zhou,[7] C. F. Chang,[3] L. H. Tjeng,[3] S. F. Lee,[8] C. F. Pai,[4] M. Hong,[2,a)] and J. Kwo[1,a)]

[1] *Department of Physics, National Tsing Hua University, Hsinchu 300044, Taiwan*

[2] *Graduate Institute of Applied Physics and Department of Physics, National Taiwan University, Taipei 106216, Taiwan*

[3] *Max Planck Institute for Chemical Physics of Solids, Nöthnitzer Strasse 40, 01187 Dresden, Germany*

[4] *Department of Materials Science and Engineering, National Taiwan University, Taipei 106216, Taiwan*

[5] *Materials Analysis Division, Taiwan Semiconductor Research Institute, National Applied Research Laboratories, Hsinchu 300091, Taiwan*

[6] *National Synchrotron Radiation Research Center, Hsinchu 30076, Taiwan*

[7] *Institut für Ionenstrahlphysik und Materialforschung, Forschungszentrum Dresden-Rossendorf e.V., Bautzner Landstraße 128, 01328 Dresden, Germany*

[8] *Institute of Physics, Academia Sinica, Taipei 11574, Taiwan*

[#] *M. X. G., C. K. C., and Y. C. L. have made equal contributions to the work.*

[a)] *Authors to whom the correspondence should be addressed:*

*mhong@phys.ntu.edu.tw and raynien@phys.nthu.edu.tw*

**ABSTRACT**

Single-crystal europium iron garnet (EuIG) thin films epitaxially strain-grown on gadolinium gallium garnet (GGG)(100) substrates using off-axis sputtering have strain-induced perpendicular magnetic anisotropy (PMA). By varying the sputtering conditions, we have tuned the europium/iron (*Eu/Fe*) composition ratios in the films to tailor the film strains. The films exhibited an extremely smooth, particle-free surface with root-mean-square roughness as low as 0.1 nm as observed using atomic force microscopy (AFM). High-resolution x-ray diffraction (XRD) analysis and reciprocal space maps (RSM) showed in-plane epitaxial film growth, very smooth film/substrate interface, excellent film crystallinity with a small full width at half maximum (FWHM) of 0.012° in the rocking curve scans (ω scans), and an in-plane compressive strain without relaxation. In addition, spherical aberration-corrected scanning transmission electron microscopy (Cs-corrected STEM) showed an atomically abrupt interface between the EuIG film and GGG. The measured squarish out-of-plane magnetization-field hysteresis loops by vibrating sample magnetometry (VSM) in conjunction with the measurements from angle-dependent x-ray magnetic dichroism (XMCD) demonstrated the PMA in the films. We have tailored the magnetic properties of the EuIG thin films, including saturation magnetization ($M_s$) ranging from 71.91 to 124.51 emu/c.c. (increase with the (*Eu/Fe*) ratios), coercive field ($H_c$) from 27 to 157.64 Oe, and the strength of PMA field ($H_\perp$) increasing from 4.21 to 18.87 kOe with the in-plane compressive strain from -0.774 to -1.044%. We have also investigated spin transport in Pt/EuIG bi-layer structure and evaluated the real part of spin mixing conductance to be $3.48 \times 10^{14}$ $\Omega^{-1}$m$^{-2}$. We demonstrated the current-induced magnetization switching with a low critical switching current density of $3.5 \times 10^6$ A/cm$^2$, showing excellent potential for low-dissipation spintronic devices.

## I. INTRODUCTION

Rare-earth iron garnets, as magnetic insulators (MIs), have played an essential role in the development of spintronics. Previously, yttrium iron garnet (YIG) was widely utilized in spin-wave related research for the efficiency of magnetic excitation because of its small magnetization damping.[1,2] In addition to the spin dynamics, the insulating property led to the discovery of spin Hall magnetoresistance (SMR),[3] which is based on the spin Hall effect in heavy metals (HMs) with strong spin-orbit coupling (SOC) and the reflection of spin current at the interface between HM and MI. Magnetic thin films with perpendicular magnetic anisotropy (PMA) are generally favored for the scalability and stability in magneto-resistive memory devices. Therefore, an MI with a strong PMA offers a great advantage in expanding its industry applications. However, single-crystal YIG thin films grown on gadolinium gallium garnet (GGG) substrate possess in-plane magnetic anisotropy (IMA), as caused by their large shape anisotropy.[4] Both theoretical[5,6] and experimental[7,8] studies have shown that a suitable substrate may induce lattice strain to give rise to PMA in YIG; the strength, nonetheless, is still weaker than that of its counterparts of thulium iron garnet (TmIG)/GGG[9] and europium iron garnet (EuIG)/GGG.[10]

The strain-induced PMA was reported in thin films of TmIG and terbium iron garnet (TbIG), most of which were grown by pulsed laser deposition (PLD).[11-16] Wu *et al.* employed off-axis magnetron sputtering to deposit TmIG MI thin films with PMA[9,17] and studied their magnetic properties, and the spin transport properties of HM/TmIG bi-layer structures. Furthermore, PMA of TmIG has enabled the studies of breaking time-reversal symmetry in the topological insulator using interfacial exchange coupling.[18,19] Compared to TmIG films, EuIG films have shown stronger strength of

PMA field ($H_\perp$) and larger coercive field ($H_c$).[10, 20]

In this work, we utilized the off-axis magnetron sputtering to deposit single-crystal strained-EuIG films epitaxially on GGG(001). These sputtered films are of single domain with a very narrow full width at half maximum of 0.012° in the rocking curve scans, compared with 0.011° of GGG substrate, a very smooth surface (with roughness of 0.1 nm) free of particles, and an atomically sharp EuIG/GGG interface. We have varied the europium to iron (*Eu/Fe*) ratio (at.%/at.%) from 0.498 to 0.646 to produce in-plane compressive strain from -0.774 to -1.044% to tailor the magnetic properties of the EuIG films. The strain has induced PMA in our films, whose strength, as measured/ calculated using SMR, increases with the in-plane compressive and the out-of-plane tensile strain. The measurements of our EuIG films using vibrating sample magnetometry (VSM) and the X-ray magnetic circular dichroism (XMCD) also established the PMA. We fine-tuned the $H_c$ and $H_\perp$ over a much broader range than the previous work of TmIG and EuIG films, making our EuIG films for versatile applications. For example, although the SMR-induced anomalous Hall effect (AHE) in Pt/EuIG was reported previously,[10, 20] the current-induced magnetization switching has been lacking. By tuning the $H_c$ to a suitable value, and meanwhile by maintaining good squareness of the AHE loop, here we have demonstrated the current-induced magnetization switching in the Pt/EuIG bilayer structure.

## II. EXPERIMENTAL

EuIG thin films were deposited on GGG(001) substrates by RF magnetron off-axis sputtering with a ~0.4 nm/min growth rate. The substrates were cleaned with acetone, isopropanol, and deionized water sequentially in an ultrasonic bath. The substrates were then dried under nitrogen gas flow. Each of the cleaned substrates was adhered to a holder with silver paste for better thermal contact during film deposition. The pressure during the deposition ($P_{dep}$) was 3 to 8 mTorr with a mixture of Ar and $O_2$ under a flow rate of 40 and 0.7 standard cubic centimeter per minute (sccm), respectively. EuIG films were deposited at 450°C, followed by post-deposition annealing at 450°C for 10 minutes in an $O_2$ ambient. The film properties were adjusted by varying the longitudinal target-to-substrate distance ($L$) and $P_{dep}$. A schematic of the sputtering set-up was provided in Fig. S1 of the supplementary material.

Atomic force microscopy (AFM) was used to characterize the surface morphology of the films and to determine the film thickness. X-ray diffraction (XRD) using synchrotron radiation was performed at beamline BL13A and BL17B of Taiwan Light Source in National Synchrotron Radiation Research Center (NSRRC), Hsinchu, Taiwan, to study the crystallinity, the epitaxy, the film thickness, and the lattice parameters of the single-crystal EuIG films to determine in-plane and out-of-plane strains of the films. Cs-corrected scanning transmission electron microscopy (Cs-STEM) was performed to probe the atomic-scale EuIG/GGG interfacial structures. The film compositions and *Eu/Fe* ratios were determined by Rutherford backscattering spectrometry (RBS) in random-geometry measurements using a 1.7 MeV $He^+$ beam performed at Helmholtz-Zentrum Dresden Rossendorf (HZDR), Germany. The backscattered particles were detected at an angle of 170° with respect to the incoming beam direction.

The magnetization ($M_s$) and the coercive field ($H_c$) were measured using VSM at room temperature. X-ray absorption spectroscopy (XAS) and XMCD taken at the Fe $L_3$ edge using a total electron yield (TEY) mode were conducted at beamline TPS45A of Taiwan Photon Source, NSRRC, Taiwan, to measure PMA in the EuIG films. The films were magnetized before loading to the end station and were measured with $Fe_2O_3$ crystals simultaneously in another chamber for the relative energy reference.

SMR transport measurements were carried out in a Quantum Design Physical Property Measurement System (PPMS) with a rotator capable of angular-dependent MR measurements. Pt/EuIG bilayer samples were fabricated into Hall bars (650 μm × 50 μm) using photolithography and then were connected to the corresponding channels on the sample holder with copper wire and silver paste.

The current-induced magnetization switching was performed by injecting the pulsed currents along the current channel with a pulse-width of $t_{\text{pusle}} = 0.05$ s from a Keithley 2400 source meter. The magnetization was monitored by the Hall voltage and collected by a Keithley 2000 multimeter. During the current-induced magnetization switching measurement, an in-plane magnetic field ($H_x$) was applied along the current channel to break the domain wall symmetry and therefore, the spin torque can drive the domain nucleation and facilitate the domain wall propagation.[21,22]

## III. RESULTS AND DISCUSSION

### A. Film morphology, crystal structure, composition, and strain

The study on surface morphology of the EuIG thin films using AFM (Figure 1(a)) showed a root-mean-square (RMS) roughness as low as 0.1 nm. The atomically flat and particle-free film surface was essential for the subsequent growth of high-quality heterostructures. A high-angle annular-dark-field (HAADF) image taken along [010] direction in Fig. 1(b) shows the epitaxial growth of the EuIG(001) film on GGG(001) substrate with a nearly perfected EuIG/GGG interface, having no defects and dislocations. The intensity contrast between the EuIG film and the GGG was caused by the mass difference of *Eu/Fe* and *Gd/Ga* in EuIG and GGG, respectively.

Figure 2(a) shows the XRD measurements along the surface normal (001) for selected EuIG films in varying $L$ from 5 cm (*Sample A*, $P_{dep}$= 3 mTorr), 6 cm (*Sample B*, $P_{dep}$= 3 mTorr), 8 cm (*Sample C*, $P_{dep}$= 3 mTorr), to 10 cm (*Sample D*, $P_{dep}$= 3 mTorr), and a higher $P_{dep}$ of 8 mTorr (*Sample E*, $L$ = 6 cm). Note that $L$ is described in Fig. S1. The out-of-plane (OOP) lattice constants ($a_{\perp,f,strained}$) of *Sample A, B, C, D,* and *E* were determined to be 12.561 Å, 12.584 Å, 12.594 Å, 12.600 Å, and 12.624 Å, respectively, with the corresponding film thickness from AFM being 33.2, 33.4, 26.4, 25.4, and 41 nm, respectively, and from XRD 29.6, 33.2, 26.9, 26.7, and 41 nm, respectively. Note that all the samples are epitaxially strained on the GGG substrate. The reported bulk lattice constant of GGG is 12.382 Å, and that of EuIG is 12.497 Å.[23,24] The red-dashed line shown in Fig. 2(a) denotes the peak position of bulk EuIG. The peak positions of the EuIG(004) reflection were at lower angles than that of the GGG(004) reflection in all samples, indicating a larger $a_{\perp,f,strained}$ of EuIG films than the lattice constant of GGG substrate and the compressively strained growth of the EuIG films.

By assuming the same elastic constants in bulk EuIG of different *Eu/Fe* ratios, the relaxed lattice constants of samples having different *Eu/Fe* ratios were determined by the equation $\frac{c_{11}}{c_{11}+2c_{12}}\frac{a_s-a_{\perp,f,strained}}{a_{f,relaxed}} = \frac{a_s-a_{f,relaxed}}{a_{f,relaxed}}$, where elastic constant $c_{11}$ and $c_{12}$ are $25.10\times10^{11}$ dyne/cm$^2$ and $10.70\times10^{11}$ dyne/cm$^2$, respectively. $a_s$ denotes the lattice constant of substrate, $a_{f,relaxed}$ denotes the lattice constant of relaxed film, and $a_{\perp,f,strained}$ denotes the out-of-plane lattice constant of strained film. The relaxed lattice constants of *Sample A, B, C, D,* and *E* were determined to be 12.479 Å, 12.491 Å, 12.496 Å, 12.500 Å, and 12.513 Å, respectively.

The in-plane strains ($\varepsilon_\parallel$) and out-of-plane strains ($\varepsilon_\perp$) were calculated by the equations of $\varepsilon_\parallel = \frac{a_s - a_{f,relaxed}}{a_{f,relaxed}}$, where $a_s$ (in-plane lattice constant of substrate) is the same as the in-plane lattice constant of the strained film ($a_{\parallel,f,strained}$), because of fully strained growth of the EuIG films on GGG substrate, and $\varepsilon_\perp = \frac{a_{\perp,f,strained} - a_{f,relaxed}}{a_{f,relaxed}}$, respectively. The measured OOP lattice constants ($a_{\perp,f,strained}$), the calculated relaxed constants, the in-plane and out-of-plane strains for *Sample A, B, C, D, and E* are listed in Table 1. As the *Eu/Fe* ratio in EuIG film increases, the in-plane and out-of-plane strain increases, respectively.

High crystalline quality in the sputtered single-crystal films is evidenced from the observation of clear and pronounced thickness fringes around the EuIG(004) peak and the narrow full width at half maximum (FWHM) of the rocking curve scans (ω scans) of 0.012° comparable to that (0.011°) of GGG substrate in Fig. 2(b).

The sample with the largest $a_{\perp,f,strained}$ (*Sample E*) was chosen for reciprocal space map (RSM) measurement. Figure 2(c) shows an RSM around the (204) off-normal reflections of both EuIG and GGG, which is plotted as a function of $Q_x$ and $Q_z$. The $Q_x$ positions of EuIG(204) and GGG (204) were both located at 1.616 nm$^{-1}$, indicating that the EuIG film is fully strained on GGG along the *in-plane* direction. The epitaxial orientation relationship between the EuIG film and GGG substrate was determined to be EuIG(001)[100]//GGG(001)[100] according to the RSM results from the off-normal diffraction peaks. Moreover, the off-normal phi-scan of another sample grown at L=6 cm with the same growth condition as *Sample B* was taken for confirming the film to be of a single domain. From the $Q_z$ position of the EuIG film and GGG, the $a_{\perp,f,strained}$ of the EuIG was determined to be 12.627 Å, consistent with the value (12.624 Å) determined from the normal scan along the EuIG(001) direction as shown in Fig. 2(a).

Knowing the different lattice constants of the film and substrate, $a_{\perp,f,strained}$'s elongation indicates that the film is under an *in-plane* compressive strain. Furthermore, the gradual increase in $a_{\perp,f,strained}$ of the strained epitaxial films indicates that under $P_{dep}$ = 3 mTorr during the sputtering, the *Eu/Fe* ratio in the films varies with the longitudinal target-to-substrate distance in the sputtering chamber, *L*.

We employed RBS to measure the chemical compositions of *Sample A, B, C,* and *D*, with their *Eu/Fe* ratios being 0.498, 0.532, 0.578, and 0.586, respectively. Note that the stoichiometric *Eu/Fe* ratio of EuIG ($Eu_3Fe_5O_{12}$) is 0.6. Figure 3(a) shows the $a_{\perp,f,strained}$ values versus the *Eu/Fe* ratios, indicating a linear increase of $a_{\perp,f,strained}$ from 12.561 to 12.600 Å with *Eu/Fe* from 0.498 to 0.586. The fitting details of RBS data are displayed in Fig. S2 of the supplementary material. The *Eu/Fe* ratio of *Sample E* is 0.646, as extrapolated from Fig. 3(a). The ionic radius of $Eu^{3+}$ (0.107 nm) is larger than that of $Fe^{3+}$ (0.064 nm),[25] attributing to the measured experimental results. For the studied sputtered Fe-rich EuIG films, the excessive Fe would occupy the Eu at the dodecahedral site, leading to a decrease in $a_{\perp,f,strained}$.[26, 27]

The variation of *Tm/Fe* ratio in TmIG thin films has affected the magnetic properties.[9, 17] In this work, we tuned *Eu/Fe* ratios to attain the desirable film strains, and thus to manipulate the magnetic properties. Figure 3(b) shows the in-plane and out-of-plane strains, as listed in Table 1, versus the *Eu/Fe* ratios for *Sample A, B, C,* and *D*. In our sputtered EuIG films, the film strain increases with the *Eu/Fe* ratio.

We have performed XRD scans on another set of films, denoted as *Sample A', B', C',* and *D'*, which were prepared using the same sputtering conditions as *Sample A, B, C,* and *D*. The XRD results were very similar between the two sets of the samples, as shown in Fig. 3(a). This demonstrates the reproducibility of our sputtering deposition.

## B. Magnetic properties

### (a) PMA measured from M-H loops using VSM

Figures 4(a) to (d) show a sequence of *out-of-plane* M-H hysteresis loops of *Sample A, B, C,* and *D* with the *Eu/Fe* ratios of 0.498, 0.532. 0.578, and 0.586, respectively. $M_s$ values for these samples were measured to be 71.91, 95.12, 120.52, and 124.51 emu/c.c., respectively. Figure 4(e) shows an *in-plane* M-H loop for *Sample D*.

Note that the $M_s$ value for the bulk EuIG is 93 emu/c.c., and those for the pulsed laser deposited (PLD) EuIG films are 110 emu/c.c. (on GGG(111)) and 120 emu/c.c. (on GGG(100)). [20] The $M_s$ values for the PLD EuIG films on GGG(111) by another research group[10] ranged from 71.56 to 74.17 emu/c.c.. In the Fe-rich film, excess $Fe^{3+}$ occupies the $Eu^{3+}$ site and reduces the total moment because of the smaller magnetic moment of $Fe^{3+}$ than that of $Eu^{3+}$.

The 100% squareness of the M-H loops demonstrates the attainment of PMA in the EuIG films. The measured magnetic coercive fields ($H_c$) of the sputtered single-crystal EuIG films in this work increased with $a_{\perp,f,strained}$ (thus the *Eu/Fe* ratio), from 27 to 158 Oe but decreased to 80 Oe for *Sample D* with an $a_{\perp,f,strained}$ of 12.6 Å. In comparison, the $H_c$ values for the PLD EuIG films on GGG(100) were 400 Oe (56 nm thick) and 100 Oe (26 nm), while those for the films on GGG(111) were 20 Oe (56 nm) and < 5Oe (26 nm). [20] The $H_c$ for the 38 nm thick PLD EuIG film on GGG(001) by another research group[10] was ~750 Oe.

The values of $M_s$ and $H_c$ versus *Eu/Fe* ratios are plotted in Figure 4(f).

### (b) PMA of EuIG FILMs via XAS/XMCD

To further probe the PMA of our sputtered EuIG films, we have carried out the XMCD measurements, with a schematic shown in Fig. 5(a), to examine whether the magnetic moment has an *in-plane* component in our films. A sample 12.52 nm in thickness was grown under the same condition as *Sample B*, followed by deposition of a 2 nm thick Ti layer by e-beam evaporation. Figure 5(b) shows the angle-dependent XAS and XMCD of Fe $L_3$ edge. The negative peak at a photon energy of 709 eV is from the Fe ions located at the tetrahedral sites (Fe$_{tet}$). The two positive peaks next to the negative peak are from the Fe ions located at octahedral sites (Fe$_{oct}$). The results indicate that Fe$_{oct}$ is antiferromagnetically coupled to Fe$_{tet}$.[28, 29] The relative XMCD intensity is largest at the normal incidence (zero incident angle) configuration; namely, the moment parallel to the incident light. The XMCD intensity ($I_{XMCD}$) decreases with the increasing incident angle. Figure 5(c) plots the XMCD intensity as a function of incident angle, and the data can be fitted very well by a simple cosine function. This XMCD result demonstrates the PMA of the EuIG film, which has no in-plane component.

### (c) Strength of PMA via SMR

To quantitatively evaluate the strength of PMA of the sputtered EuIG films, we extracted the $H_\perp$, utilizing electrical transport measurements on sputtered Pt (3nm thick)/EuIG films. The samples were then patterned into a Hall bar geometry with 650 μm in length and 50 μm in width using standard photolithography.

We have performed the SMR measurements of a series of Pt/EuIG bilayers of various *Eu/Fe* ratios with the *in-plane* magnetic field applied transverse to the current. The longitudinal magnetoresistance ratio $\triangle R_{xx}/R_{xx}(0)$ versus *in-plane* magnetic field ($H_y$) along the y-axis is plotted in Figs. 6(a) to (d), where $\Delta R_{xx}=R_{xx}(H)-R_{xx}(0)$. According to the SMR theory, when the magnetization is aligned in the y-direction by the applied field, the resistance reaches the minimum because of less spin absorption at the interface. Thus, by applying the relationship $H_{in-sat} = H_\perp - 4\pi M_s$, where $H_{in-sat}$ stands for the *in-plane* saturation field. The PMA strength ($H_\perp$) values of *Sample A"* (9.7 nm in thickness), *B"* (10.5 nm), *C"* (12.5 nm), and *D"* (13 nm) were measured/ calculated to be 4.21, 10.95, 15.97, and 18.87 kOe, respectively. Note that *Sample A", B", C"*, and *D"* were prepared in the same sputtered conditions as *Sample A, B, C,* and *D*, respectively. We, therefore, expect that the *Eu/Fe* ratios, the $a_{\perp,f,strained}$ values, and the film strains are similar between these two sets of samples. The larger *in-plane* compressive strain ($\varepsilon_\parallel$) (or equivalently the out-of-plane strain ($\varepsilon_\perp$)) as caused by the increase of Eu in the film has attributed to the enhanced PMA strength ($H_\perp$), as shown in Fig. 6(e) and 6(f).

In comparison, the $H_\perp$ values attained in the sputtered single-crystal TmIG films (24.5 nm in thickness) by Wu *et al.* ranged from 1.43 to 2.44 kOe. The $H_\perp$ value for the PLD EuIG film on GGG(100) was 1.88 kOe, and those for the PLD EuIG films by Ortiz *et al.* ranged from 4.13 kOe (56 nm) to 32.91 kOe (4 nm).

**(d) Spin mixing conductance and current-induced magnetization switching**

The transport and current-induced switching measurements were carried out on the Pt (3 nm)/EuIG (9.7 nm) (*Sample A"*) sample with the aforementioned Hall bar geometry. According to the SMR theory,[30] the transverse Hall resistivity ($\rho_{trans}$) in an HM/MI bilayer can be expressed as follows:

$$\rho_{trans} = \Delta\rho_1 m_x m_y + \Delta\rho_2 m_z \qquad (1)$$

where $m_i$ denotes the *i*-component of the unit magnetization of EuIG. From $\Delta\rho_1$ and $\Delta\rho_2$, we arrive at the following relations:

$$\frac{\Delta\rho_1}{\rho} = \theta_{SH}^2 \frac{\lambda}{d_N} \frac{2\lambda G_r \tanh^2 \frac{d_N}{2\lambda}}{\sigma + 2\lambda G_r \coth \frac{d_N}{\lambda}} \qquad (2)$$

$$\frac{\Delta\rho_2}{\rho} \approx \theta_{SH}^2 \frac{\lambda}{d_N} \frac{2\lambda \sigma G_i \tanh^2 \frac{d_N}{2\lambda}}{(\sigma + 2\lambda G_r \coth \frac{d_N}{\lambda})^2} \qquad (3)$$

where $\rho$, $\sigma$, $d_N$, $\lambda$, $\theta_{SH}$, $G_r$ and $G_i$ represent Pt longitudinal resistivity, conductivity, the thickness of the metal layer, spin diffusion length, spin Hall angle, the real and the imaginary parts of spin mixing conductance, respectively. First, we measured the SMR-induced anomalous Hall signal with the *out-of-plane* magnetic field at room temperature. Good squareness of AHE loop with the AHE loop coercivity ($H_c^{AHE}$) of 65 Oe is clearly shown in Fig. 7 (b). From the amplitude of the AHE signal, we then obtained $\Delta\rho_2 = 9.63\times10^{-4}$ μΩ-cm. Next, we measured the *in-plane* angular-dependent transverse resistance displayed in Fig. 7(a), and $\Delta\rho_1 = 4.94 \times 10^{-2}$ μΩ-cm was extracted from the fitting result. $\rho = 67.5$ μΩ-cm was measured in the same Hall bar. We assumed $\lambda = 1.4$ nm and $\theta_{SH} = 0.08$.[31, 32] The values of $G_r$ and $G_i$ are calculated to be $3.48 \times 10^{14}$ Ω$^{-1}$m$^{-2}$ and $1.13 \times 10^{13}$ Ω$^{-1}$m$^{-2}$, respectively. Note that the previous work by Rosenberg *et al.*[20] reported the lower bound of $G_i$ to be $5.4 \times 10^{12}$ Ω$^{-1}$m$^{-2}$ of Pt/EuIG/GGG(001) according to their AHE measurement and the $G_r$ from the reference of Pt/TmIG. Here, by measuring both SMR-induced AHE and *in-plane* angular dependent SMR, we directly obtained the precise $G_r$ and $G_i$ values of the Pt/EuIG interface.

The current-induced magnetization switching was demonstrated on the same device with the transport measurement. Figure 7(c) represents the switching results with the *in-plane* field of ±60 Oe. This external *in-plane* field breaks the switching symmetry and overcomes the interfacial Dzyaloshinskii–Moriya interaction (DMI) effective field. Compared to the AHE loop, the magnetization of EuIG is fully switched by the pulsed current-induced spin-orbit torque from Pt. The critical current density ($J_c$) is $3.47 \times 10^6$ A/cm$^2$, which is lower than the values obtained in Pt/TmIG ($J_c = 1.8 \times 10^7$, $6.0 \times 10^6$ A/cm$^2$) by Avci *et al.,* and is comparable to the value of $J_c = 2.5 \times 10^6$ A/cm$^2$ by Wu *et al.*[9] The lower critical current density could be attributed to the higher $G_r$, which gives rise to the higher efficiency of spin transmission at the interface. This has been achieved by the relatively small $H_c$ and $M_s$ that can be precisely adjusted by manipulating the *Eu/Fe*.

## IV. CONCLUSION

Single-crystal EuIG thin films with excellent crystallinity and smooth surface were epitaxially grown on GGG(001) using the off-axis sputtering. Fully strained EuIG epi films on GGG have shown PMA, which was established by both squarish out-of-plane M-H loop and angle-dependent XMCD measurement. Clear SMR-induced AHE loops with good squareness were measured through the spin transport measurement, enabling the calculation of the $H_\perp$ values (PMA strength) of the insulating EuIG films. The PMA strength was well correlated with the measure/calculated film strain, which was tuned with the *Eu/Fe* ratio in the film. For EuIG films possessing suitably low $H_c$, we have fabricated a Pt/EuIG structure to attain a low switching current density for the current-induced PMA magnetization switching, suggesting the potential for constructing low-dissipation spintronic devices. Also, the sputtering technique is advantageous as it can be scaled up for industrial applications.

**ACKNOWLEDGMENTS**

The work was supported by the Ministry of Science and Technology of Taiwan under Grant No. 105-2112-M-007-014-MY3, 110-2112-M-002-036, and 110-2622-8-002-014. We thank the TEM technical services of NTU Consortium of Electron Microscopy Key Technology. We also acknowledge the support from the Max Planck-POSTECH-Hsinchu Center for Complex Phase Materials.

**FIGURES**

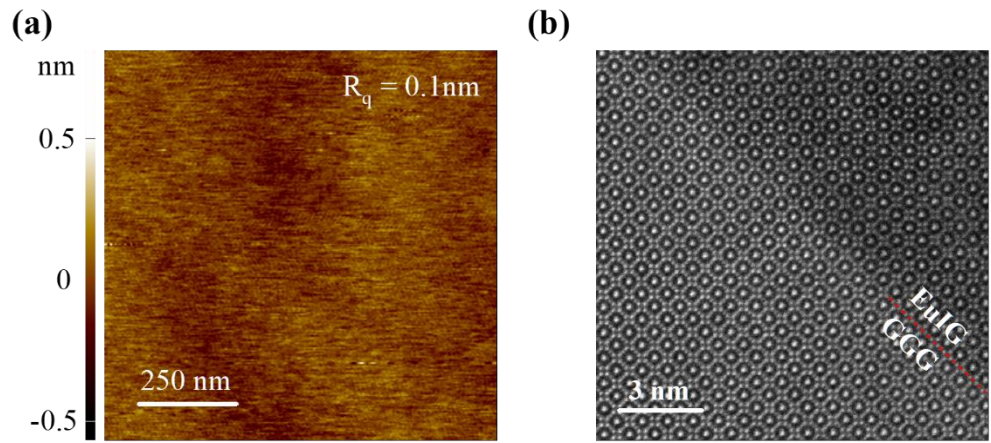

FIG. 1. (a) Surface morphology image of a 24 nm thick EuIG film (*Eu/Fe* = 0.532) grown on GGG(100). $R_q$ stands for RMS roughness. (b) Cs-corrected STEM HAADF cross-sectional image at the interface of EuIG and GGG (*Eu/Fe* = 0.532) with zone axis [010]. The red dash line indicates the interface between EuIG and GGG.

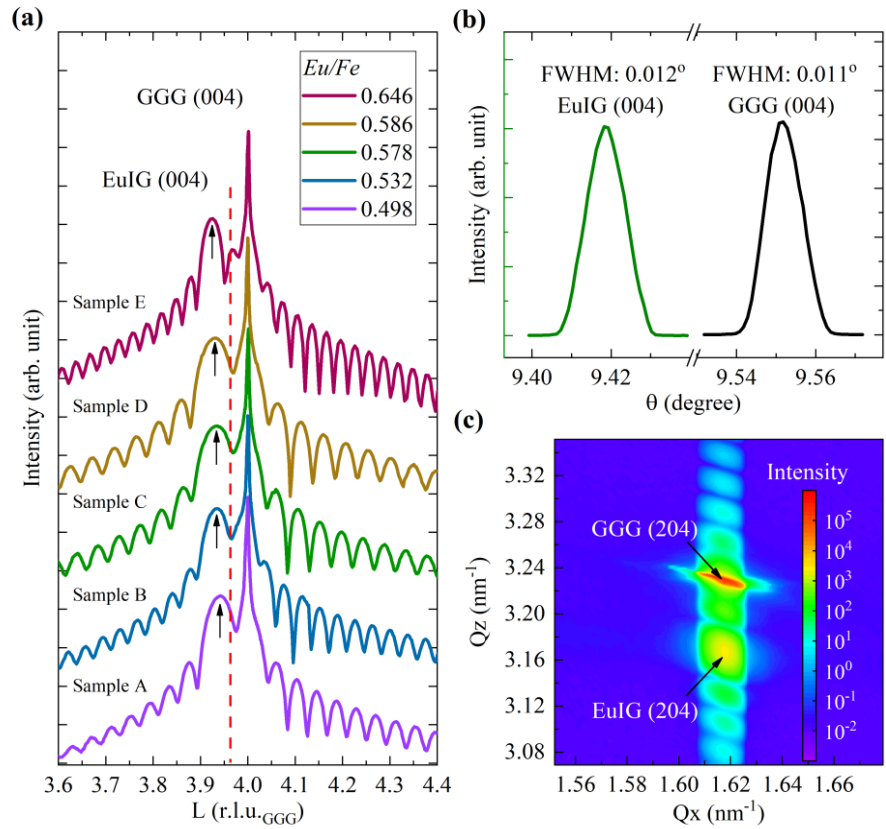

FIG. 2. (a) XRD normal scans for EuIG films grown on GGG(100) substrate with different *Eu/Fe* ratios, showing clear Laue oscillations. The arrows indicate the position of EuIG(004) peak. The red dashed line indicates the position of bulk EuIG for reference. (b) FWHM of rocking curve scans of EuIG(004) and GGG(004) is 0.012° and 0.011°, respectively. (c) RSM for EuIG(204) and GGG(204) diffraction of a 41 nm thick EuIG film (*Sample E*).

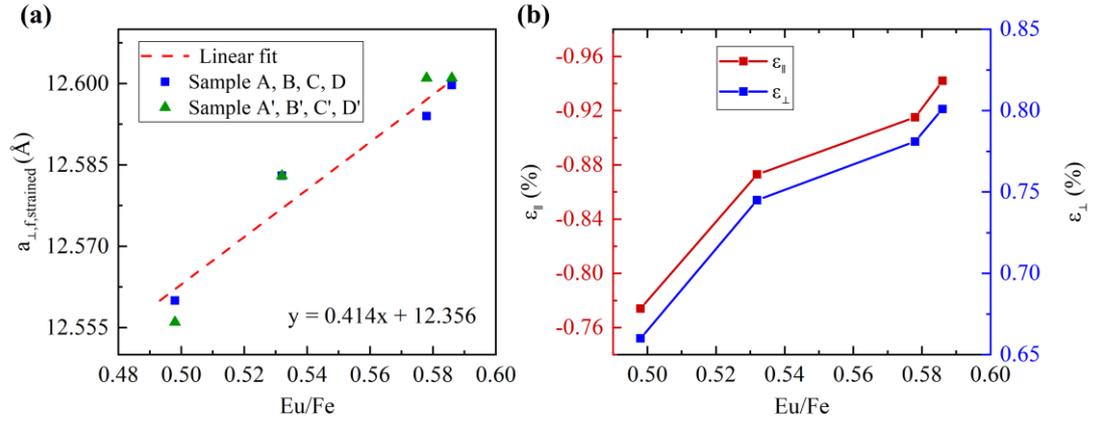

FIG. 3. (a) Out-of-plane (OOP) lattice constants ($a_{\perp,\text{f,strained}}$) of EuIG films versus *Eu/Fe* values. The *Eu/Fe* ratio is at. % of Eu divided by at. % of Fe in the sample. The red dash line is a linear fit to the four data points. (b) Plots of in-plane strain ($\varepsilon_{\parallel}$) and out-of-plane strain ($\varepsilon_{\perp}$) versus *Eu/Fe* ratio.

| Sample | Relaxed $a_{f,relaxed}$ (Å) (calculated) | Strained $a_{\perp,f,strained}$ (Å) (measured) | $\varepsilon_{\parallel}$ (%) | $\varepsilon_{\perp}$ (%) |
|---|---|---|---|---|
| A | 12.4786 | 12.561 | -0.774 | 0.660 |
| B | 12.4910 | 12.584 | -0.873 | 0.745 |
| C | 12.4964 | 12.594 | -0.915 | 0.781 |
| D | 12.4997 | 12.600 | -0.942 | 0.802 |
| E | 12.5126 | 12.624 | -1.044 | 0.890 |

Table 1. Structural parameters and strains for epitaxially EuIG/GGG(001)

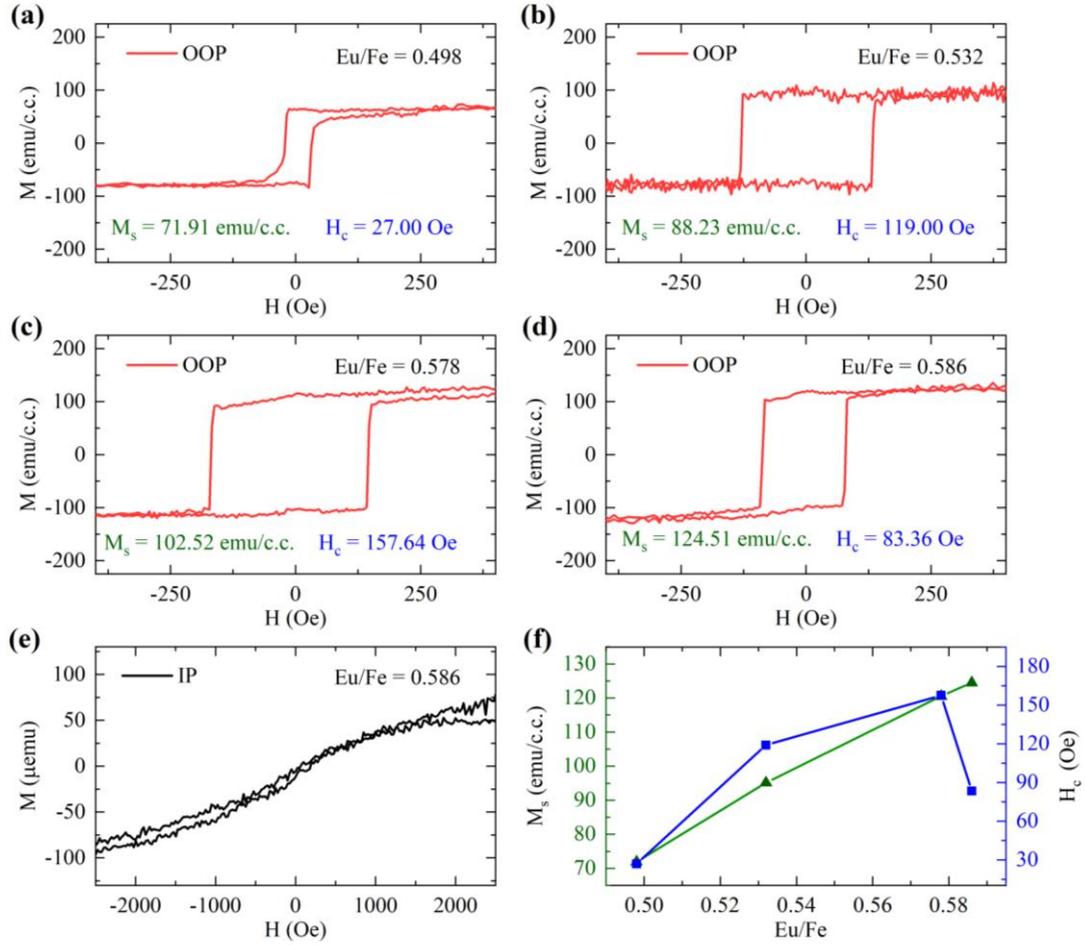

FIG. 4. *Out-of-plane* M-H loops for samples with *Eu/Fe* of (a) 0.498, (b) 0.532, (c) 0.578, and (d) 0.586. The liner background from the GGG substrate was subtracted. $M_S$ was calculated and marked at the bottom of each panel. (e) An *in-plane* M-H loop for *Eu/Fe* = 0.586 sample. Note that its linear background was not subtracted because it has not reached magnetic saturation in this field range, and the y-axis refers to the total magnetization. (f) $M_s$ and $H_c$ versus *Eu/Fe* ratios.

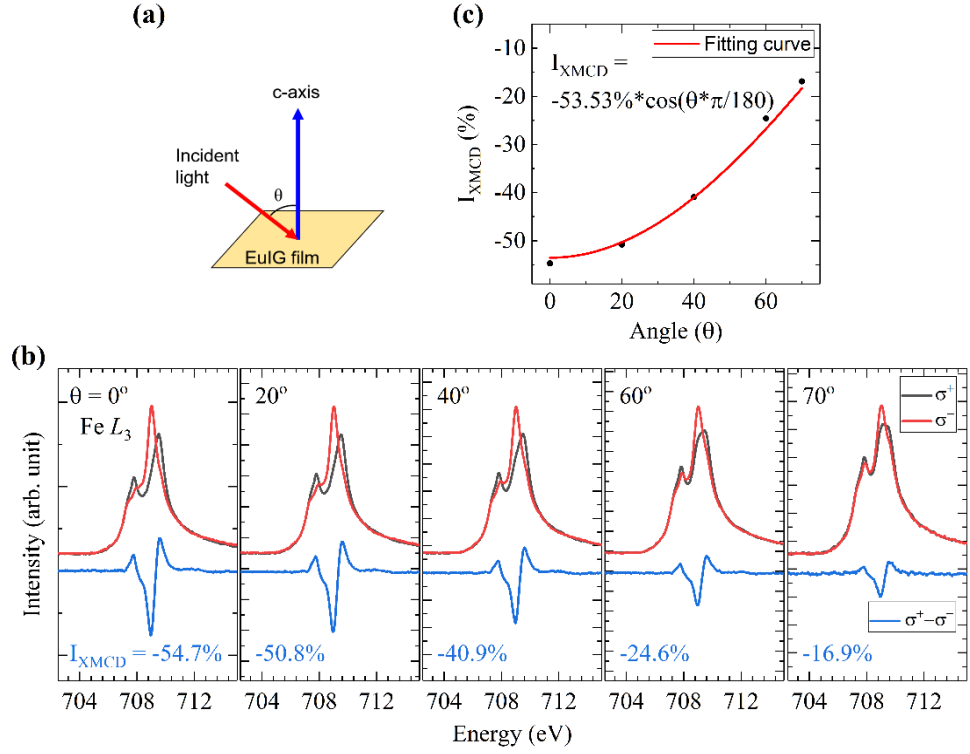

FIG. 5. Angle-dependent XAS and XMCD spectra of the Fe $L_3$ edge on the EuIG film measured at 300 K. (a) illustrates the incident angle's geometry configuration to the surface normal. (b) is the angle-dependent XMCD spectra taken at Fe $L_3$ edge, $\sigma+$ and $\sigma-$ denote the two XAS spectra taken at oppositely polarized light. The relative $I_{XMCD}$ of Fe$_{tet}$ is calculated from the height of XMCD divided by the height of XAS of Fe $L_3$. (c) is the summary of $I_{XMCD}$ results in (b), and the red line is a cosine fitting to the data.

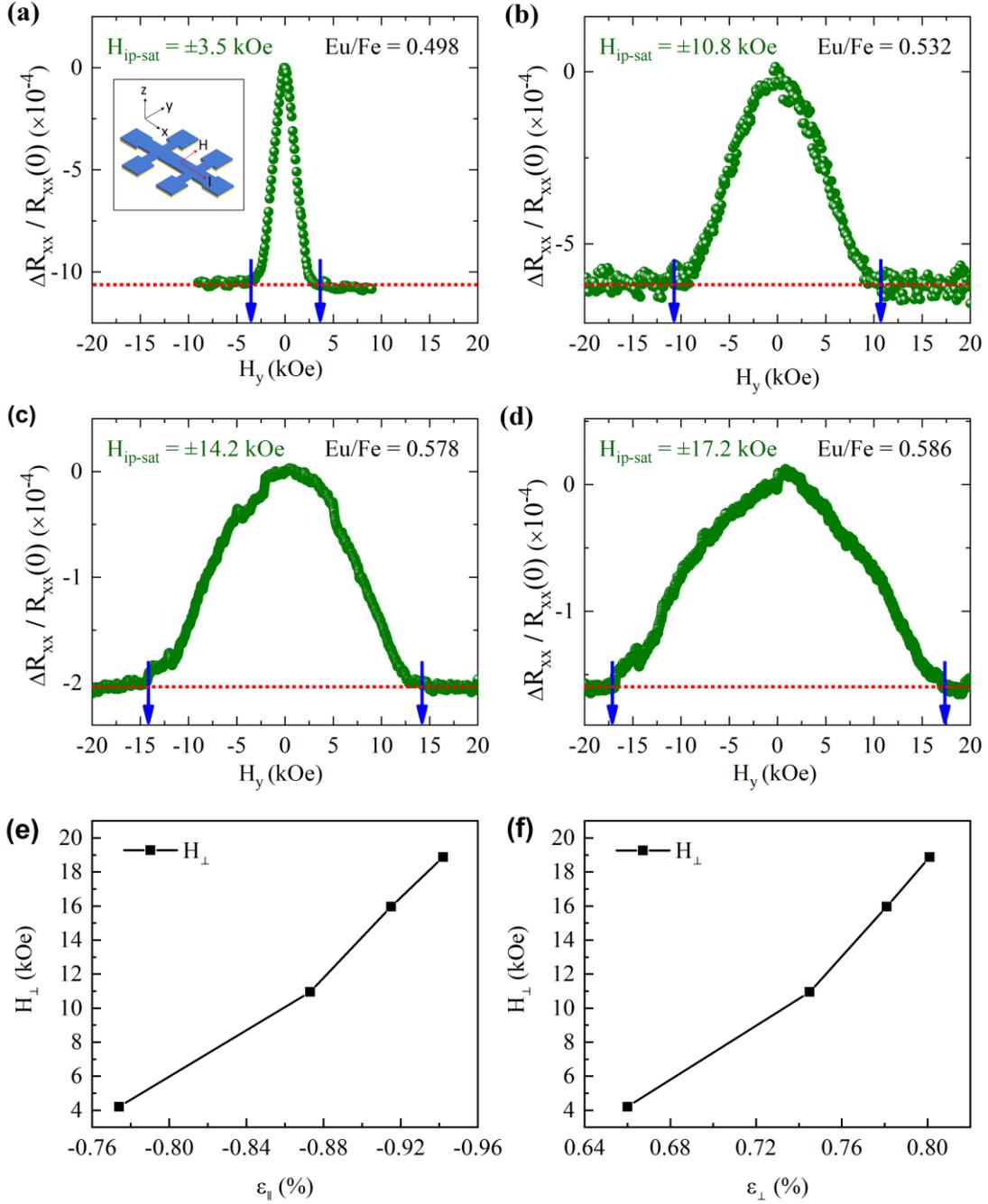

FIG. 6. SMR measurement of Pt/EuIG bilayer structures with the *in-plane* field transverse to the current as indicated in the inset of (a). (a) to (d) represent the results of longitudinal magnetoresistance ratio ($\Delta R_{xx}/R_{xx}$) as a function of *in-plane* magnetic field ($H_y$) for the EuIG films with *Eu/Fe* of 0.498, 0.532, 0.578, and 0.586, respectively. The blue arrows mark the $H_{ip\text{-}sat}$ of each sample with the values denoted at the top of each panel. (e) Plot of $H_\perp$ versus in-plane strain ($\varepsilon_\parallel$). (f) Plot of $H_\perp$ versus out-of-plane

strain ($\varepsilon_\perp$)

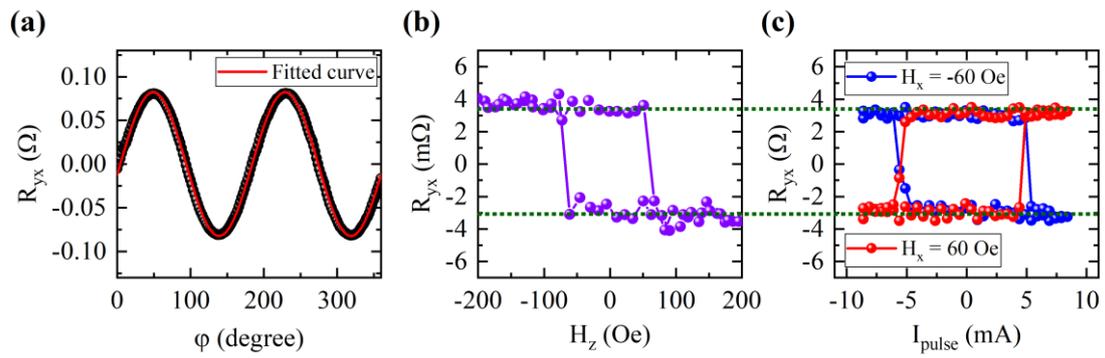

FIG. 7. (a) *In-plane* angle-dependent SMR for the Pt/EuIG (*Sample A''*, *Eu/Fe* = 0.498) sample with 10 kOe applied field. (b) and (c) show the AHE measurement and current-induced switching measurement, respectively.

## Supplementary information S1:

## Off-axis sputtering configuration

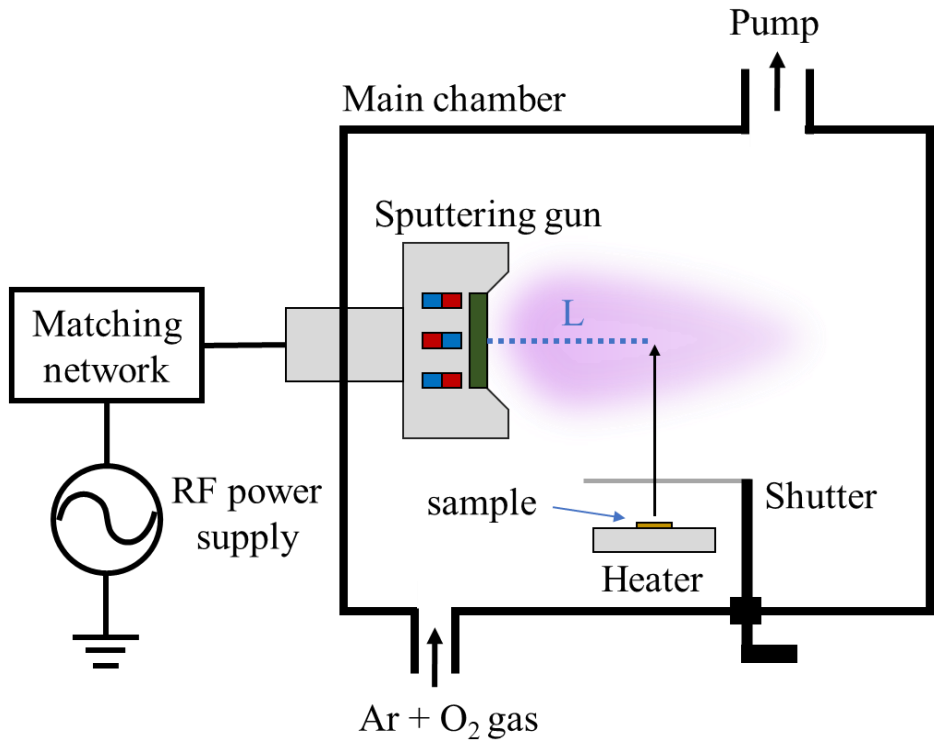

FIG. S1. A schematic of the off-axis RF magnetron sputtering set-up. The base pressure in the chamber was < $3 \times 10^{-7}$ torr. The film deposition took place with the substrates being located at various L (cm).

# Supplementary information S2:

# Compositional analysis using RBS

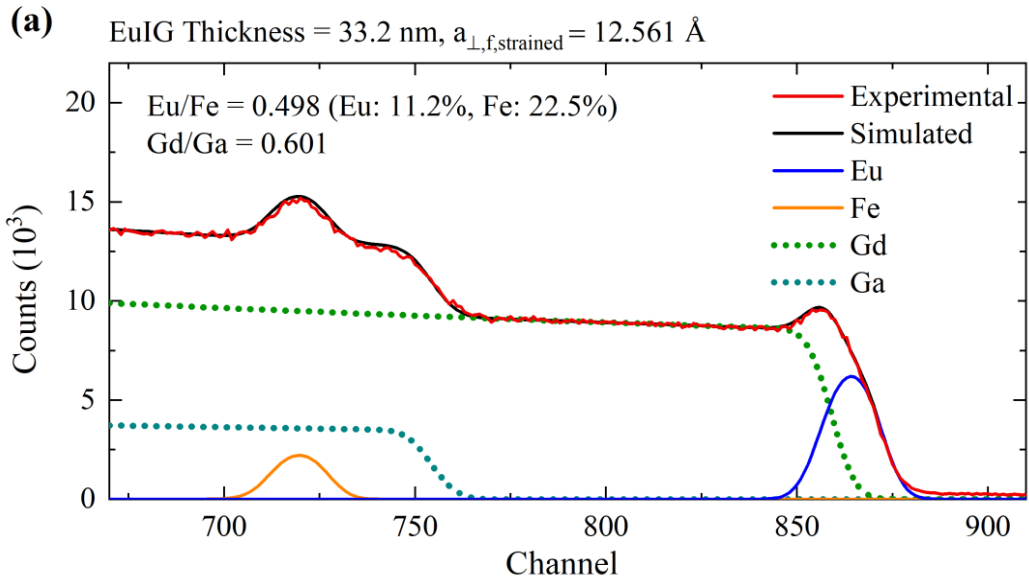

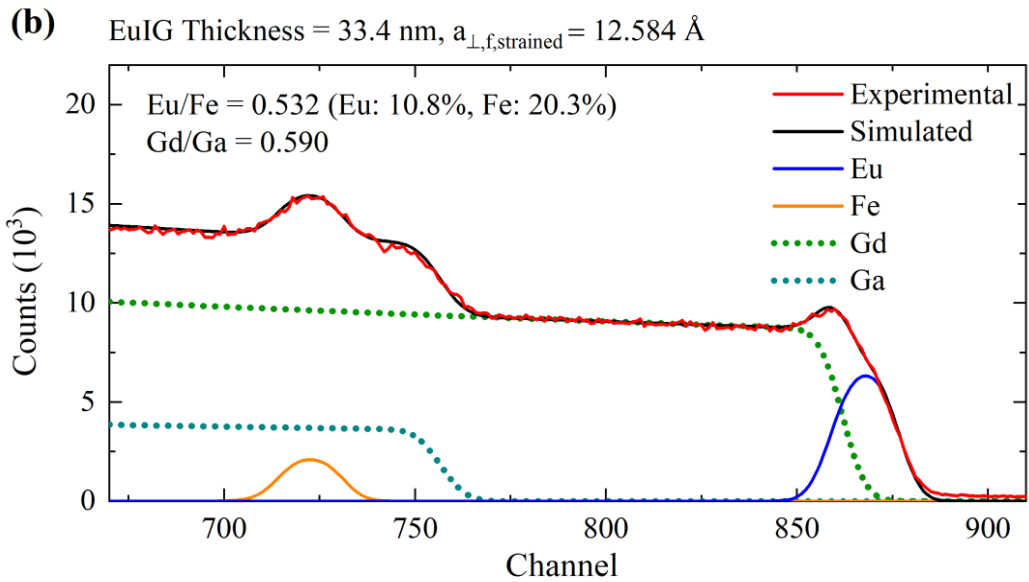

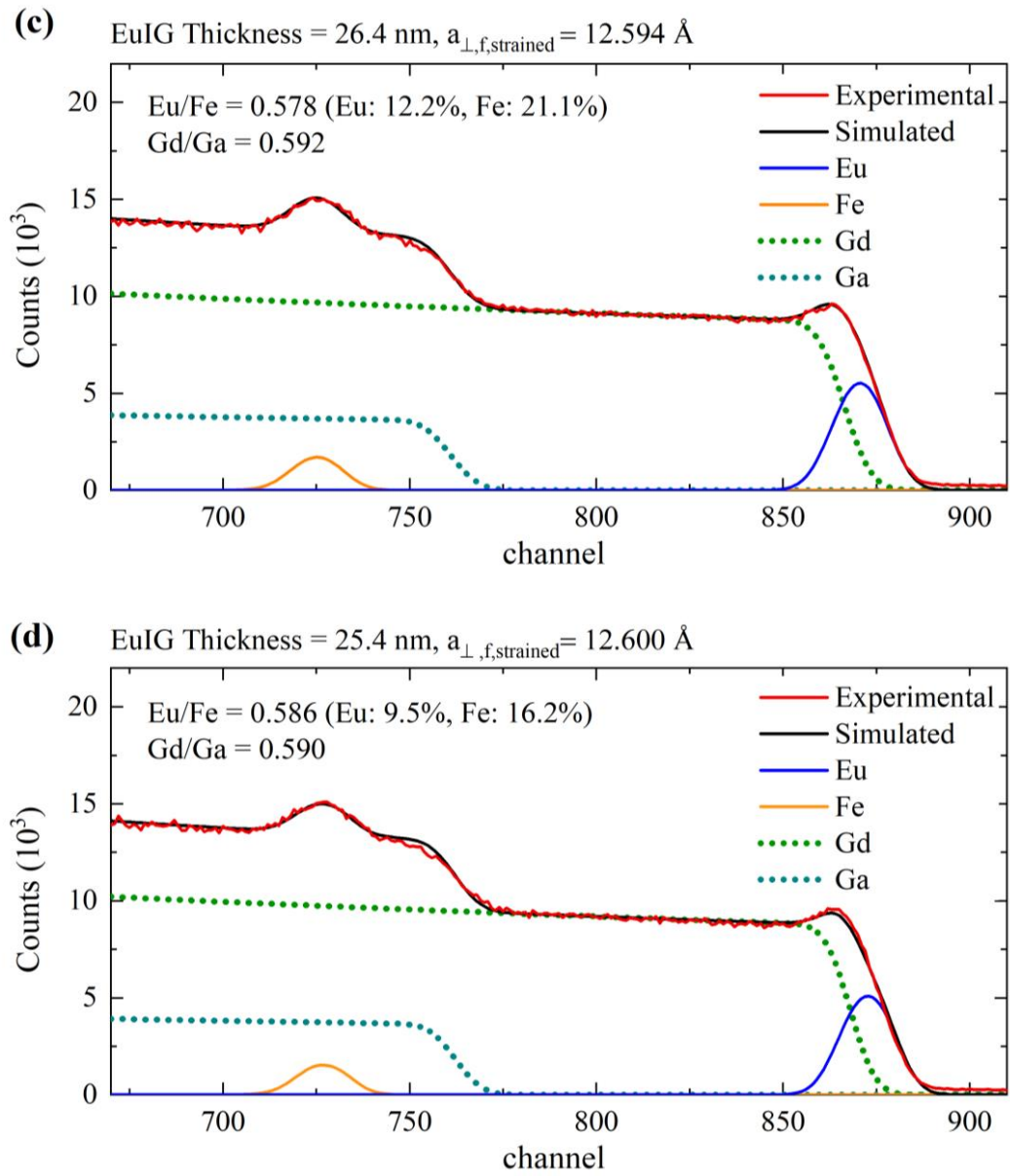

FIG. S2. (a) to (d) showing RBS composition analysis results for samples with different $a_{\perp,\text{f,strained}}$. We used SIMNRA to fit the experimental curves. The Gd/Ga ratios of the substrate were also fitted in each sample, as shown in each panel.